\documentclass[a4paper]{article}

\usepackage{newlfont}
\usepackage{amsfonts}
\usepackage{amssymb}
\usepackage{amsmath}
\usepackage{latexsym}
\usepackage[all]{xy}

\def\sT{\textsf{T}}
\def\sP{\textsf P}
\def\sV{\textsf V}
\def\sv{\textsf v}
\def\R{\mathbb{R}}
\def\A{\mathbf{A}}
\def\I{\mathbf{I}}
\def\Z{\mathbf{Z}}

\def\W{\mathbf{W}}
\def\d{\textsf{d}}
\def\<{\langle} 
\def\>{\rangle} 
\def\({\left(}
\def\){\right)}
\def\p{\mathbf p}
\newtheorem{prop}{Proposition}
\newtheorem{thm}{Theorem}
\newenvironment{pf}{{\noindent{\it Proof. }}}{\hfill \rule{2mm}{2.5mm}\medskip}
\date{ }
\title{AV-differential geometry and Newtonian mechanics\thanks{Research
supported by the Polish Ministry of Scientific Research and
Information Technology under the grant No. 2 P03A 036 25.}}

\author{Katarzyna Grabowska \\
{\tt konieczn@fuw.edu.pl}\\
Pawe\l\ Urba\'nski \\
{\tt urbanski@fuw.edu.pl} \\
Physics Department, University of Warsaw\\
Ho\.za 69, 00-681 Warszawa }

\begin{document}

\maketitle


\begin{abstract}
A frame independent formulation
of analytical mechanics in the Newtonian space-time is presented.
The differential geometry of affine values (AV-differential geometry) i.e., the differential
geometry in which affine bundles replace vector bundles and
sections of one dimensional affine bundles replace functions on
manifolds, is used. Lagrangian and hamiltonian generating objects,
together with the Le\-gendre transformation independent on
inertial frame are constructed.



\end{abstract}

\section{Introduction}
Mathematical formulation of analytical mechanics is usually based
on objects that have vector character. So is the case of the most
of mathematical physics. We use tangent vectors as infinitesimal
configurations, cotangent vectors as momenta, we describe dynamics
using forms (symplectic form) and multivectors (Poisson bracket)
and finally we use an algebra of smooth functions. However, there
are cases where we find difficulties while working with
vector-like objects. For example, in the analytical mechanics of
charged particles we have a problem of gauge dependence of
lagrangians. In Newtonian mechanics there is a strong dependence
on inertial frame, both in lagrangian and hamiltonian formulation.
In the mechanics of non-autonomous system we are forced to choose
a reference vector field on the space-time that fulfills certain
conditions or we cannot write the dynamics at all. In all those
cases the traditional language of differential geometry seems to
introduce too much mathematical structure. In other words, there
is to much structure with comparison to what is really needed to
define and describe the behavior of the system. As a consequence
we have to put in an additional information to the system such as
gauge or reference frame.

As it is well known,  Newton's equations do not depend on the
inertial frame chosen. Therefore, a geometric formulation of
Newtonian mechanics in Newtonian space-time is possible. On the
other hand, analytical mechanics in Newtonian space-time is not
possible in a standard framework. Different lagrangians and
different hamiltonians are used for different inertial frames. The
same is true for Hamilton-Jacobi theory and Schr\"odinger wave
mechanics. Frame independence of the Lagrangian formulation of
Newtonian dyna\-mics  can be achieved by increasing the dimension
of the configuration space of the particle.  The four dimensional
space-time of general relativity is replaced by the five
dimensional ma\-nifold (as in the Kaluza theory) \cite{WMT1},
\cite{WMT2}, \cite{DB}. An alternate approach is proposed in the
present note. The four dimensional space-time is used as the
configuration space.  The phase space is no longer a cotangent
bundle and not even a vector bundle.  It is an affine bundle
modelled on the cotangent bundle of the space-time manifold. The
Lagrangian is a section of an affine line bundle over the tangent
bundle of the space-time manifold. The proper geometric tools are
provided by the {\it geometry of affine values} (AV- differential
geometry). We call the geometry of affine values the differential
geometry that is built using sections of one-dimensional affine
bundle over the manifold instead of functions on the manifold. The
affine bundle we use is equipped with the fiber action of the
group $(\R,+)$, so we can add reals to elements of fibres and real
functions to sections, but there is no distinguished "zero
section". Those elements of the geometry of affine values that are
needed in the Newtonian mechanics are described in section 4.1, an
extended presentation of the theory can be found in
\cite{GGU1,GGU}. In fact the geometry of affine values appeared
much earlier in works of W.M. Tulczyjew  (see e.g. \cite{TUZ}) and
has been  applied to the description of the dynamics of charged
particles in \cite{TU}.
    Affine phase bundle for analytical mechanics in
the Newtonian space-time was discussed in papers of Tulczyjew
\cite{WMT1} and Pidello \cite{P}. Some problems concernig
analytical mechanics on affine bundles are discussed in \cite{MMS}
and \cite{SMM}.

\section{Newtonian space-time}

    The {\it Newtonian space-time} (some authors prefer to call it Galilean space-time,
    but we follow the notation of Benenti and Tulczyjew) is a system $(N,\tau, g)$ where $N$ is a
four-dimensio\-nal affine space with the model vector space $V$, $\tau$ is a
non-zero element of $V^\ast$ and $g\colon E_0\rightarrow E_0^\ast$ represents an
Euclidean metric on $E_0=\ker\tau$. The elements of the space $N$ represent events.
The time elapsed between two events is measured by $\tau$:
    $$\Delta t(x,x')=\<\tau,x-x'\>.$$
    The distance between two simultaneous events is measured by $g$:
    $$d(x,x')=\sqrt{\<g(x-x'),x-x'\>}.$$
    The space-time $N$ is fibrated over the time $T=N\slash E_0 $
which is one-dimensional affine space modelled on $\R$. By $\eta$ we will
denote the canonical projection
        $$\eta\colon N\longrightarrow T, $$
    by $\imath$ the canonical embedding
    $$\imath: E_0 \longrightarrow V ,$$
    and by $\imath^\ast$ the dual projection
    $$\imath^\ast: V^\ast \longrightarrow E_0^\ast.$$
    By means of $\imath$ and $\imath^\ast$ we can define a contravariant tensor $g'$ on
$V^\ast$:
    $$g'=\imath\circ g^{-1}\circ\imath^\ast.$$
    The kernel of $g'$ is a one-dimensional subspace  of $V^\ast$ spanned by $\tau$.

Let $E_1$ be an affine subspace of $V$ defined by the equation $\<\tau,v\>=1$. The
model vector space for this subspace is $E_0$. An element of $E_1$ can represent
velocity of a particle. The affine structure of $N$ allows us to associate to an
element $u$ of $E_1$ the family of inertial observers that move in the space-time
with the constant velocity $u$. This way we can interpret an element of $E_1$ as an
inertial reference frame. For a fixed inertial frame  $u$, we define the space  $Q$
of world lines of all inertial observers. It is the quotient affine space $N/\{u\}$.
The space-time $N$ becomes the product of affine spaces
        $$ N = Q\times T  .
                                                                $$ 
    The model vector space for $Q$ is the quotient vector space $V/\{u\}$ that can be
identified with $E_0$. The corresponding canonical projection is
    $$\imath_u \colon  V\rightarrow E_0\colon  v\longmapsto
\imath_u(v)=v-\<\tau,v\>u $$
    and the splitting $V=E_0\times \R$ is given by
    $$V\ni v\longmapsto (\imath_u(v),\<\tau,v\>)\in E_0\times\R.$$
    The dual splitting is given by
    $$V^\ast \ni p\longmapsto (\imath^\ast(p), \<p,u\>)\in E_0^\ast\times\R.$$

    The tangent bundle $\sT N$ we identify with the product $N\times V$ and the
subbundle  $\sV N$  of vectors  vertical with respect to the
projection on time, with $N\times E_0$. Consequently,  the  bundle
$\sV^1N$ of infinitesimal configurations (positions and
velocities) of particles moving in the space-time $N$ is
identified with $N\times E_1$. When the inertial frame $u$ is
chosen, $E_1$ is identified with $E_0$ and $\sV^1 N$ is identified
with $\sV N$.

The vector dual $\sV^\ast N$ for $\sV N$ is a quotient bundle of $N\times V^\ast$ by
the one-dimensional subbundle $N\times\<\tau\>$. We can identify it  with
$N\times  E_0^\ast$. Using the inertial frame we can make it a subbundle of $\sT^\ast N$.

\section{Analytical mechanics in a fixed inertial frame}

In the following section we will present the analytical mechanics of one
particle in the Newtonian space-time in the fixed inertial frame $u$.
Before we start working on physics we shall recall basic constructions and
facts about generating objects for lagrangian submanifolds in $\sT^\ast
M$. (In the homogeneous formulation of the dynamics we will be using more
general generating objects than just a function on the manifold $M$.) The
details can be found in \cite{WMT3} and \cite{LM}. Later we concentrate on
the inhomogeneous formulation of the analytical mechanics of one particle,
suitable for trajectories parameterized by the time, then we pass to the
homogeneous one. The homogeneous formulation accepts all
parameterizations.

\subsection{Generating objects for lagrangian submanifolds in the cotangent bundle}
\label{gener}

In many cases the dynamics of a mechanical system is obtained as an
inverse image by a symplectomorphism of a certain lagrangian submanifold.
It can be lagrangian submanifold of the cotangent bundle to the phase
space generated by a hamiltonian or the lagrangian submanifold of the
cotangent bundle to the space of infinitesimal configurations generated by
a lagrangian. In some cases however one needs more general generating
object than just a function on a manifold.

The most general generating object for a lagrangian submanifold of a
cotangent bundle is a family of functions i.e. a function on the total
space of a fibration over the manifold. In the following we will recall
some definitions and constructions that will be used later in the context
of hamiltonian and lagrangian dynamics.

Let $\rho: N\rightarrow M$ be a differential fibration and $F:N\rightarrow
\R$ be a smooth function. The pair $(F, \rho)$ will be called a familly of
functions in a sense that it is a familly of functions on the fibers of
$\rho$ parameterized by points of $M$. We will need the following
definition: The set
$$S(F,\rho)=\{n\in N:\; \forall v\in\sV_nN\;
\langle\d F(n),v\rangle=0\}$$ is called {\it the critical set} of
the family $(F,\rho)$.

The process of generating a lagrangian submanifold can be
described in two equivalent ways:

\noindent{\bf 1.} The function $F$ generates the lagrangian
submanifold $\d F(N)\subset\sT^\ast N$. Then we use the canonical
projection $\tilde\rho: \sV^\circ N\rightarrow \sT^\ast M$, where
$\sV^\circ N$ is the anihilator of the vertical bundle $\sV N$,
to obtain a subset $L$ of $\sT M$:
$$L=\tilde\rho(\sV^\circ N\cap\d F(N)).$$
In another words we apply the symplectic reduction with respect to
a coisotropic submanifold $\sV^\circ N$ to $\d F(N)$. We have a
theorem:

\begin{thm}[\cite{LM}]
If $V^\circ N$ and $\d F(N)$ have clear intersection than the set
$L$ is an immersed lagrangian submanifold of $\sT^\ast M$.
\end{thm}

The proof and other details of the construction can be found in \cite{LM}.

\noindent{\bf 2.} The generated set $L$ is the image of the critical set
$S(F,\rho)$ by the mapping
$$\kappa: S(F,\rho)\longrightarrow \sT^\ast M$$
defined as follows
$$\forall v\in \sT_mM\;\;\;\forall w\in \sT_nN: \sT\rho(w)=v\;\;\;
\langle\kappa(n), v\rangle=\langle\d F(n), w\rangle.$$ The conditions for
$F$ to be a generating object of a lagrangian submanifold are formulated
in a language of Hessian which may be more familiar for people working in
analytical mechanics. In the following we will concentrate on the second
approach.

Let $n$ be a critical point of the family $(F,\rho)$, let $\mathcal O$ be
a neighbourhood of $(0,0)$ in $\R^2$ and $\chi:\mathcal O\rightarrow N$ be
a smooth mapping such that $\chi(0,0)=n$ and
$\rho(\chi(s,t))=\rho(\chi(0,t))$ for all $s$. It means that the parameter
$s$ changes along the fibre of $\rho$. It is easy to show that
$$\frac{\partial^2}{\partial t\partial s}_{|t=0,s=0}F\circ\chi$$
depends only on the vectors tangent to the curves $s\mapsto\chi(s,0)$,
$t\mapsto \chi(0,t)$ in points $s=0$, $t=0$ respectively. Since it is
always possible to construct $\chi$ for a given pair of vectors we can
define the mapping
$$H(F,\rho, n): \sV_nN\times\sT_nN\longrightarrow\R,\quad
H(F,\rho,n)(v,w)=\frac{\partial^2}{\partial t\partial
s}_{|t=0,s=0}F\circ\chi.$$  One can show that $H(F,\rho,n)$ is linear in
both arguments. Moreover if $v$ and $w$ are both vertical it is symmetric.
The mapping $H(F,\rho,n)$ is called a {\it Hessian of the family
$(F,\rho)$ in the point $n$}.

We have the following definition: A family $(F,\rho)$ is called a
{\it Morse family} if the rank of the Hessian is maximal in every
point of the critical set.

\begin{thm} Te set generated by a Morse family is an immersed
lagrangian submanifold of $\sT^\ast M$.
\end{thm}

Let us look at the mechanical example: \vskip10pt

\noindent {\bf Example:} Let $Q$ be a space of configurations of
an autonomous mechanical system with first-order lagrangian $L:\sT
Q\rightarrow \R$. Let $M=\sT^\ast Q$ denote the phase space. We
take $N$ to be $\sT^\ast Q\times_Q\sT Q$ and $\rho: N\ni
(p,v)\longmapsto p\in M$. The critical set of the family
$$F(p,v)=L(v)-\langle p,v\rangle$$
is given by
$$S(F,\rho)=\{(p,v): \d_\sV L(v)=p\}.$$
By $\d_\sV$ we denote the vertical differential, i.e. the differential
along the fibre of $\sT Q\rightarrow Q$. In local coordinates the matrix
of Hessian assumes the form
$$H(F,\rho, n)=\left[\begin{array}{c}
\vspace{5pt}\frac{\partial^2L}{\partial x^i\partial v^j} \\
\vspace{5pt}\frac{\partial^2L}{\partial v^i\partial v^j} \\
\delta_{ij}
\end{array}\right]$$
and is clearly of the maximal rank equal $\dim Q$. Therefore $(F,\rho)$ is
a Morse family. It generates a lagrangian submanifold of $\sT^\ast\sT^\ast
Q$. The inverse image of the submanifold by the symplectomorphism
$\beta_Q$ associated to the symplectic form on $\sT^\ast Q$ is the
dynamics of the system. \vskip10pt

In some cases a generating object can be simplified. Suppose that
$\rho$ is a composition of two projections
$\rho=\rho_2\circ\rho_1$:
$$\xymatrix{
N\ar[r]^{\rho_1}  & P \ar[r]^{\rho_2} & M }
$$
and the critical set $S(F,\rho_1)$ is contained in an image of some
section $\sigma: P\rightarrow N$. In another words $\rho_1$ establishes
one-to-one correspondence between $S(F,\rho_1)$ and $\rho_1(S(F,\rho_1))$.
In this situation the family $(F,\rho)$ can be simplified. The same set
$L$ is generated by the family $F_1: P\rightarrow \R$,
$F_1(p)=F(\sigma(p))$.\vskip10pt

\noindent{\bf Example.} Let us consider the family from the
previous example: $F(v,p)=L(v)-\langle p,v\rangle$. The critical
set $S(f,\rho)$ is given by the equation
$$\frac{\partial L}{\partial v^i}=p_i.$$
The condition for $S(F,\rho)$ to be  locally an image of the section is
given by the implicit function theorem. In case of regular lagrangian, i.e
if the matrix $\frac{\partial^2 L}{\partial v^i\partial v^j}$ is
invertible we can locally have one-to one correspondence between
velocities and momenta. We can therefore locally simplify the family $F$
obtaining hamiltonian function. We call a lagrangian hyperregular if there
is a global diffeomofphism between $\sT M$ and $\sT^\ast M$ given by the
equation
$$\frac{\partial L}{\partial v^i}=p_i.$$
In this case family $(F,\rho)$ can be globally reduced to the hamiltonian
function.

\subsection{Inhomogeneous dynamics described in a fixed  inertial frame}
\label{inhom}

    Let $u\in E_1$ represent an inertial frame.  For a fixed time $t\in T$, the
phase space for a particle with mass $m$ with respect to the inertial frame $u$ is
$\sT^\ast N_t\simeq N_t\times E_0^\ast$, where $N_t=\eta^{-1}(t)$. The collection
of phase spaces form a phase bundle $\sV^\ast N\simeq N\times E_0^\ast$. Phase
space trajectories of the system are solutions of the well-known equations of
motion:
    \begin{equation}\label{eqmotion}\dot p =-\d_s\varphi(x)\qquad\dot
x=g^{-1}(\frac{p}{m})+u, \end{equation}
    where $(x,p,\dot x,\dot p)\in \sV^1\sV^\ast N \subset\sT\sV^\ast N\simeq
N\times E_0^{\ast}\times V\times E_0^\ast$ and $\varphi\colon N\rightarrow\R$
is a potential. Subscript ${}_s$ in $\d_{s}$ means that we differentiate only
in spatial directions i.e. the directions vertical with respect to the
projection on time, therefore $\d_{s}\varphi(x)\in E_0^\ast$. The equations
define a vector field on $\sV^\ast N$ with values in $\sV^1\sV^\ast N$, i.e.
a section of the bundle $\sV^1\sV^\ast N\rightarrow\sV^\ast N$. The image of
the vector field (\ref{eqmotion}) we will call {\it the inhomogeneous
dynamics} and denote by $D_{i,u}$.
    It can be generated directly by the \textbf{lagrangian}
\begin{equation}\label{inhomlag}
    \ell_{i,u}\colon \sV^1 N \rightarrow \R \colon (x,w)\mapsto
\frac{m}2\<g(w-u),w-u\>-\varphi(x).
 \end{equation}

 With this lagrangian we associate the Legendre mapping
 \begin{equation}\label{legi}
\begin{array}{rcl}
  \cal L_{i,u} &\colon &\sV^1 N \rightarrow \sV^\ast N \\
   &\colon &(x,w) \mapsto
     g\circ \imath_u(v),
\end{array}
\end{equation}
    i.e. the vertical derivative of $\ell_{i,u}$ with respect to the projection $\sV
    N \rightarrow N$.

    The procedure of generating the inhomogeneous dynamics from the lagrangian
(\ref{inhomlag}) is as follows.  The image of the vertical
derivative $\d_s \ell_{i,u}$ is a submanifold of
    \[\sV^\ast \sV^1 N \simeq N\times E_1\times E_0^\ast \times E_0^\ast  \]
which is canonically isomorphic (as an affine space) to
    \[\sV^1\sV^\ast N \simeq N\times  E_0^\ast \times E_1 \times E_0^\ast . \]
    This isomorphism we obtain by a reduction of the canonical diffeomorphism
    $$\alpha_M \colon \sT\sT^\ast M\rightarrow \sT^\ast \sT M $$
    valid for any differential manifold $M$
(for the definition of $\alpha_{M}$ see \cite{WMT3}), which for
the affine space $N$ assumes the form
    \[\alpha_N(x,a, v, b) = (x,v, b, a).\label{alf}\]
After the reduction, we get
    \[\alpha_N^1 \colon N\times  E_0^\ast \times E_1 \times E_0^\ast\rightarrow
    N\times E_1\times E_0^\ast \times E_0^\ast. \]
Using $\alpha_N^1$ we can obtain $D_{i,u}$ from $\d_s \ell_{i,u}$
by taking an inverse-image:
    \[D_{i,u}=(\alpha_N^1)^{-1}(\d_s \ell_{i,u}(\sV^1N)).\]

    The dynamics $D_{i,u}$ cannot be generated directly from a hamiltonian
by means of the canonical Poisson structure  on $\sV^\ast N$,
which is the reduced cano\-nical Poisson (symplectic) structure of
$\sT^\ast N$. In the coordinates adapted to the structure of the
bundle $(t,x^i, p_i)$ the Poisson bi-vector is given by
$$\Lambda=\partial p_i\wedge \partial x^i.$$
    Symplectic leaves for this Poisson structure are cotangent bundles $\sT^\ast
N_t$, where $N_t =\eta^{-1}(t)$. It follows that every hamiltonian vector
field is  vertical with respect to the projection on time. However, using
reference frame $u$, we can generate first the vertical part of the dynamics
(\ref{eqmotion}), i.e the equations

    \begin{equation}\label{veqmotion}\dot p =-\d_s\varphi(x)\qquad\dot
x=g^{-1}(\frac{p}{m}), \end{equation}
    and add the reference vector field $u$.
{\bf The hamiltonian function} for the problem reads
$$h_{i,u}(x,p)=\frac{1}{2m}\<p,g^{-1}(p)\>+\varphi(x),$$
where $(x,p)\in\sV^\ast N$.

 The system (\ref{veqmotion}) can be
generated also from lagrangian function defined on $\sV N$ by the formula:
    \begin{equation}
\ell_{i,u}(x,w)=\frac{m}2\<g(w),w\>-\varphi(x).
    \end{equation}
    We identify the fiber over $t$ of $\sV N$ with $\sT N_t$ and use the
standard procedure to generate a submanifold $D_{u,t}$ in $\sT\sT^\ast N_t$.
The collection of these submanifolds give us the system (\ref{veqmotion}).

    The dynamics (\ref{eqmotion}) and the generating procedures depend
strongly on the choice of the reference frame. In particular, the relation
velocity-momenta is frame-dependent which means that we have to redefine the
phase manifold for the particle to obtain frame-independent dynamics.
    Also the hamiltonian formulation will be possible if we replace the
canonical Poisson tensor by a more adapted object.

\subsection{Homogeneous dynamics described in a  fixed inertial frame}

    In the homogeneous formulation of the dynamics infinitesimal
configurations are pairs $(x,v)\in N\times V^+ \subset N\times V\simeq \sT N$
where $V^+$ is an open set of vectors  such that $\langle \tau,v \rangle
>0$.

The homogeneous lagrangian is an extension by homogeneity of the $\ell_{i,u}$
from (\ref{inhom}) and is given by the formula:
\begin{equation}
\ell_{h, u}(x,v)=\frac{m}{2\<\tau,v\>}\<g(\imath_u(v)), \imath_u(v)\>-
\<\tau,v\>\varphi(x).
\end{equation}

    This choice guaranties that the action calculated for a piece of the world
line, which is one dimensional oriented submanifold of the
space-time, does not depend on its parametrization. However, we
still have to use the fixed inertial frame $u$.

The image of the differential of $\ell_{h,u}$ is a lagrangian submanifold of
$\sT^\ast\sT N\simeq N\times V\times V^\ast\times V^\ast$. An element
$(x,v,a_x,a_v)$ is in the image of $\d \ell_{h,u}$ if it satisfies
the following equations
\begin{equation}\label{homd}
\left\{
\begin{array}{l}
v\in V^+, \\
a_x=-\<\tau,v\>\d\varphi(x), \\
a_v=\frac{m}{\<\tau,v\>}\imath_u^\ast\circ g\circ \imath_u(v)-
 \frac{m}{2\<\tau,v\>^2}\<g(\imath_u(v)),\imath_u(v)\>\tau
 -\varphi(x)\tau.
\end{array}\right.
\end{equation}
The image of $\d \ell_{h,u}(N\times V^+)$ by the mapping $\alpha_N^{-1}$ is a
lagrangian submanifold of $\sT\sT^\ast N$. This submanifold we will call {\it
the homogeneous dynamics} and denote by $D_{h,u}$. An element $(x,p,\dot
x,\dot p)$ of $\sT\sT^\ast N\simeq N\times V^\ast\times V\times V^\ast$ is in
$D_{h,u}$ if
\begin{equation}\label{homd2}
\left\{
\begin{array}{l}
\dot x=v, \\
\dot p=-\<\tau,v\>\d\varphi(x), \\
p=\frac{m}{\<\tau,v\>}i_u^\ast\circ g\circ \imath_u(v)-
  \frac{m}{2\<\tau,v\>^2}\<g(\imath_u(v)),\imath_u(v)\>\tau
 -\varphi(x)\tau
\end{array}\right.
\end{equation}
for some $v\in V^+$, i.e. $\<\tau,v\> > 0$. We observe that $D_{h,u}$ does
not project on the whole $\sT^\ast N$, but $(x,p)$ must satisfy the following
equation:
\begin{equation}
\frac{1}{2m}\<p,g'(p)\>+\<p,u\>+\varphi(x)=0. \label{kmu}
\end{equation}
The equation (\ref{kmu}) is the analog of the mass-shell equation
$p^2=m^2$ in the relativistic mechanics. Since there is the
difference in signature of $g'$ between the Newtonian and the
relativistic case, we obtain here a paraboloid of constant mass
instead of relativistic hyperboloid. The mass-shell will be
denoted by $K_{m,u}$.

  It is possible  to  generate the dynamics  $D_{h,u}$ directly  by a
generalized hamiltonian system. The hamiltonian generating object (see
\cite{TU2}) is the family

\begin{equation} \label{fam1}\xymatrix@C=2cm{
N\times V^\ast\times V^+ \ar[r]^-{-H_{h,u}} \ar[d]_\zeta& \R\\
N\times V^\ast & },\end{equation}
    where
\begin{equation}
H_{h,u}(x,p,v)=\<p,v\>-\ell_{h,u}(x,v)\in \R.
\end{equation}

    This family can be simplified. The fibration $\zeta$ can be represented
as a composition $\zeta''\circ \zeta'$, where
    \[\zeta'\colon N\times V^\ast\times V^+ \rightarrow N\times V^\ast\times\R_+
    \colon (x,p,v)\mapsto (x,p,<\tau,v>), \]
and
     \[\zeta''\colon N\times V^\ast\times \R_+ \rightarrow N\times V^\ast
    \colon (x,p,r)\mapsto (x,p). \]
    Equating to zero the derivative of $H_{h,u}$ along the fibres of $\zeta'$
    we obtain the relation

\begin{equation}
    v = \frac{<\tau,v>}{m} g^{-1}\circ \imath^\ast (p) + <\tau,v>u.
\end{equation}

    It follows that the family (\ref{fam1}) is equivalent (generates the same
object) to the reduced family
    \begin{equation} \label{fam2}\xymatrix@C=2cm{
N\times V^\ast\times \R_+ \ar[r]^-{-\widetilde{H}_{h,u}} \ar[d]_-{\zeta''}& \R\\
N\times V^\ast & },\end{equation}
    where
    \begin{equation}
\widetilde{H}_{h,u}(x,p,r)=r(\frac{1}{2m}\<p,g'(p)\>+\<p,u\>+\varphi(x)).
\end{equation}
    No further simplification is possible.

    The critical set $S(\widetilde{H}_{h,u}, \zeta''))$ is the submanifold
    \[\left\{(x,p,r)\in N\times V^\ast\times \R_+ ;\quad \frac{1}{2m}\<p,g'(p)\>
    +\<p,u\>+\varphi(x)=0\right\}\]
    and its image $\zeta''(S(\widetilde{H}_{h,u}, \zeta''))$ is the mass shell
$K_{m,u}$.

    The function $H_{h,u}$ is zero on $S(\widetilde{H}_{h,u}, \zeta'')$ and projects to
the zero function on $K_{m,u}$. However, a Dirac system with the
zero function on the constraints $K_{m,u}$ does not generate
$D_{h,u}$. The lagrangian submanifold  $\bar{D}_{h,u}\subset
\sT\sT^\ast N$ generated by this system is exactly the
characteristic distribution of $K_{m,u}$, i.e.
$$\bar{D}_{h,u}=(\sT K_{m,u})^\S$$ and does not respect the condition
$\langle\tau,v\rangle>0$. We have only $D_{h,u}\subsetneq \bar{D}_{h,u}$.

\section{The dynamics independent on inertial frame}
\subsection{Special vector and affine spaces}
    A vector space $W$ with distinguished non-zero element $v$ we will call
    a {\it special vector space}.
 A canonical example of a special vector space is
$(\R,1)$. It will be denoted by $\I$. If $A$ is an affine space
then Aff$(A,\R)$ -- the vector space of all affine functions with
real values on $A$ -- is a special vector space with distinguished
element $1_A$ being a constant function on $A$ equal to 1. The
space Aff$(A,\R)$ will be denoted by $A^\dagger$ and called a {
\it vector dual} for $A$. Having a special vector space $(W,v)$ we
can define its {\it affine dual} by choosing a subspace in
$W^\ast$ of those linear functions that take the value $1$ on $v$:
$$W^{\ddagger}=\{\varphi\in W^\ast:\quad \varphi(v)=1\}.$$
We have that
\begin{thm}[\cite{GGU}]
For $(W,v)$ and $A$ such that \emph{dim}~$W < \infty$ and
\emph{dim}~$A < \infty$
$$\begin{array}{c}
\left((W^\ddagger)^\dagger, 1_{W^\ddagger}\right)=V, \\
\left(A^\dagger\right)^\ddagger=A.
\end{array}$$
\end{thm}
An affine space modelled on a special vector space will be called a {\it
special affine space}. Similar definitions we can introduce for bundles: a
{\it special vector bundle} is a vector bundle with distinguished
non-vanishing section and a {\it special affine bundle} is an affine
bundle modelled on a special vector bundle.

\subsection{The geometry of affine values} \label{gav}

The geometry of affine values is, roughly speaking, the
differential geometry built on the set of sections of
one-dimensional special affine bundle $\zeta:\Z\rightarrow M$
modelled on $M\times \I$, instead of just functions on $M$. The
bundle $\Z$ will be called a {\it bundle of affine values}. Since
$\Z$ is modelled on $M\times\I$ we can add reals in each fiber of
$\Z$, i.e $\Z$ is  an $\R$-principal bundle. The vertical vector
field on $\Z$ which is the fundamental vector field for the action
of $\R$ will be denoted by $X_\Z$. Let us now consider an example
of a bundle of affine values: If $(A,v)$ is a  special affine
space modelled on $(W,v)$, then we have the quotient affine space
$\underline{A}=A\slash \<v\>$. The affine spaces $A$ and
$\underline{A}$ together with the canonical projection form an
example of a bundle $\A$ of affine values. The appropriate action
of $\R$ in the fibers is given by
$$ A\times\R\ni(a,r)\longmapsto (a-rv)\in A$$
and the fundamental vector field $X_\A$ is a constant vector field
equal to $v$ on $A$.

The affine analog of the cotangent bundle $\sT^\ast M$ in the
geometry of affine values is called a {\it phase bundle} and
denoted by $\sP \Z$.  We define an equivalence relation in the set
of pairs of $(m,\sigma)$, where $m\in M$ and $\sigma$ is a section
of $\Z$. We say that $(m,\sigma)$, $(m',\sigma')$ are {\it
equivalent} if $m=m'$ and $\d(\sigma-\sigma')(m)=0$, where we have
identified the difference of sections of $\Z$ with a function on
$M$. The equivalence class of $(m,\sigma)$ is denoted by
$\d\sigma(m)$. The set of equivalence classes is denoted by
$\sP\Z$ and called the {\it phase bundle} for $\Z$. It is, of
course, the bundle over $M$ with the projection
$\d\sigma(m)\mapsto m$. It is obvious that
$$\sP\zeta:\sP\Z\rightarrow M : \d\sigma(m) \mapsto m$$ is
an affine bundle modelled on the cotangent bundle $\sT^\ast
M\rightarrow M$.

The structure of $\sP\Z$ is similar to the structure of the
cotangent bundle $\sT^\ast M$. In particular on $\sP\Z$ there is a
canonical symplectic form defined as follows. Any section $\sigma$
of the bundle $\Z\rightarrow M$ gives the trivialization
$$I_\sigma:\Z\longrightarrow M\times\R,\quad I_\sigma(z)=(\zeta(z),
z-\sigma(\zeta(z))\,)$$ and further
$$I_{\d\sigma}:\sP\Z\longrightarrow \sT^\ast M,\quad
I_{\d\sigma}(\alpha)=\alpha-\d\sigma(\sP\zeta(\alpha)).$$ For two
sections $\sigma$, $\sigma'$ the mappings $I_{\d\sigma}$ i
$I_{\d\sigma'}$ differ by the translation by the affine form
$\d(\sigma-\sigma')$, i.e.
$$I_{\d\sigma'}\circ I_{\d\sigma}^{-1}: \sT^\ast M\longrightarrow
\sT^\ast M,\quad \phi_m\longmapsto\phi_m+\d(\sigma-\sigma')(m).$$
Using the well-known property of the canonical symplectic form
$\omega_M$ on $\sT^\ast M$ that {\it translation by closed forms
are symplectomorphisms} we conclude that
$I_{\d\sigma}^\ast\omega_M$ does not depend on the choice of
$\sigma$ and therefore it is a canonical symplectic form  on
$\sP\Z$. It will be denoted by $\omega_\Z$. More information about
the canonical symplectic form $\omega_\Z$ and about the structure
of $\sP\Z$ can be found in \cite{GGU} and \cite{U}.

As an example we construct a phase bundle for the bundle of affine values
built out of a special affine space $(A,v)$. In the set of all sections of the
bundle $\A$ there is a distinguished set of affine sections, since $A$ and
$\underline{A}$  are affine spaces. We observe that there are
affine representatives in every equivalence class $\d\sigma(m)$ that
differ by a constant function. Every choice of a reference point in A defines a  one-to-one correspondence between affine sections of $\A$ and affine sections of the model bundle $\W$. This correspondence projects to the bijection between corresponding phase spaces, which does not depend on the choice of the reference point.  There is also one linear representative for each element in $\sP\W$,
i.e. such an affine section that takes value $0$ at the point $0\in \underline{W}$.
The set of elements of a phase bundle can be therefore identified with a
set of pairs: point in $m$ and a linear injection from $\underline{W}$ to $W$.
Moreover, we observe that such linear injections are in one-to-one
correspondence with linear functions on $W$ such that they take value $1$
on $v$ (or the canonical vector field $X_\W$ evaluated on the function
gives $1$). The image of a linear section is a level-$0$ set of the
corresponding function. The functions that correspond to linear sections
form the affine dual $W^\ddagger$, therefore we have
\begin{equation}\label{dag}
  \sP\W\simeq \underline{W}\times W^\ddagger, \ \ \  \sP\A\simeq \underline{A}\times W^\ddagger.
\end{equation}

Let us now see what is a canonical symplectic form on $\sP\A$. For
it we take $\sigma$ -- an affine section of $\A$ and $F_\sigma$ --
a corresponding affine function on $\A$ such that it's linear part
$\d F_{\sigma}$ takes value $1$ on $v$. The image of $\sigma$ is a
level-zero set for $F_\sigma$. We see that
$$ I_\sigma: \A\longrightarrow\underline{A}\times\R,\qquad
a\longmapsto  (\underline{a},\,a-\sigma(\underline{a})),$$ and
further
$$ I_{\d\sigma}: \underline{A}\times W^{\ddagger}\longrightarrow
\sT^\ast\underline{A}\simeq
\underline{A}\times\underline{W}^\ast,\qquad (\underline{a},f)
\longmapsto (\underline{a},\, f-\d F_{\sigma}).$$ Identifying
$\sT\sT^\ast\underline{A}$ with
$\underline{A}\times\underline{W}^\ast\times\underline{W}\times\underline{W}^\ast$
we get
$$\omega_{\underline{A}}(\,(\underline{a},\varphi, x,\psi),\, (\underline{a},\varphi,
x',\psi')\,)= \psi'(x)-\psi(x')$$ and therefore the same
expression for $\omega_\A$ reads
$$\omega_{\A}(\,(\underline{a},f, x,\psi),\, (\underline{a},f,
x',\psi')\,)= \psi'(x)-\psi(x'),$$ where $\sT\sP\A$ is identified
with $\underline{A}\times W^\ddagger\times\underline{W}\times
\underline{W}^\ast$.

\subsection{Frame independent lagrangian}

Now, we will collect all the homogeneous lagrangians for all inertial frames \label{fil}
and construct for them a universal  object which does not depend on an
inertial frame. It is convenient to  treat a lagrangian as a section of the
trivial bundle $N\times V\times \R\rightarrow N\times V$ rather than as a
function.

For two reference frames $u$ and $u'$, we have the  difference
    $$\ell_{h,u}(x,v)-\ell_{h,u'}(x,v)=m\<g(u'-u),\imath_{\frac{u'+u}{2}}(v)\>.$$
Let us denote  $\imath_{\frac{u'+u}{2}}^\ast g(u'-u)$ by $\sigma(u',u)$. With
this notation

\begin{equation}
  \ell_{h,u}(x,v)-\ell_{h,u'}(x,v)=m\<\sigma(u',u),v\>. \label{lla}
\end{equation}

For $\sigma$ we have the following equalities
\begin{equation}
\sigma(u',u)=-\sigma(u,u'), \label{sa}
\end{equation}
\begin{equation}
\sigma(u'',u')+\sigma(u',u)=\sigma(u'',u). \label{sb}
\end{equation}

In the $E_1\times N\times V\times\R$, we introduce the following relation:
\begin{equation}
(u,x,v,r)\sim (u',x',v',r')\quad\Longleftrightarrow\quad\left\{
\begin{array}{l}
x=x',\\
v=v',\\
r=r'+m\<\sigma(u',u),v\>.
\end{array}\right.\label{row}
\end{equation}
>From (\ref{sa}) we obtain that $\sim$ is symmetric and reflexive,
from (\ref{sb}) that it is transitive, therefore it is an
equivalence relation. Since the relation does not affect $N$ at
all, it is obvious that in the set of equivalence classes we have
a cartesian product structure $N\times W$. In $W$ we distinguish
two elements: $w_0=[u,0,0]$ and $w_1=[u,0,-1]$,
$$w_0=\{(u,0,0):\,\, u\in E_1\},\qquad w_1=\{(u,0,-1):\,\, u\in E_1\},$$
and two natural operations:
$$+: W\times W\rightarrow W\qquad \circ:\R\times W\rightarrow W$$
\begin{equation}
\begin{array}{l}
[u,v,r]+[u',v',r']= \\
\quad=[\frac{u+u'}{2},\, v+v',\,
r+r'+m(\langle\sigma(u,\frac{u+u'}{2}),v\rangle+
\langle\sigma(u',\frac{u+u'}{2}),v'\rangle)], \\
\lambda\circ[u,v,r]=[u,\lambda v, \lambda r].
\end{array}
\end{equation}
The above operations are well defined that can be checked by direct
calculation. Some more calculation one needs to show that

\begin{prop}
$(W,+,\circ)$ is a vector space with $w_0$ as the zero-vector. Moreover
$(W,w_1)$ is a special vector space such that $W\slash<w_1>\simeq V$
\end{prop}

\noindent The canonical projection $W\rightarrow V$ will be denoted by
$\zeta$.
    It follows from (\ref{lla}) that quadruples $(u,x,v,\ell_{h,u}(x,v))$ and
    $(u',x,v,\ell_{h,u'}(x,v))$ are equivalent. Consequently, frame
 dependent lagrangian defines a section $\ell_h$ over $N\times V^+$ of  the
one-dimensional special affine bundle (a bundle of affine values) $N\times
W\rightarrow N\times V$ which does not depend on the inertial frame.
 The section $\ell_h$ will be called an {\it
affine lagrangian} for the homogeneous mechanics independent on
the choice of inertial frame. In the following we show that the
bundle $N\times W\rightarrow N\times V$ carries a structure, which
can be used for generating the frame-independent dynamics. We
begin with the construction of the phase space.

\subsection{Phase space} \label{ps}

In the  frame dependent formulation of the dynamics,  the phase space for the
massive particle is $\sT^\ast N\simeq N\times V^\ast$. For each frame $u$ we
have the Legendre mapping
\begin{equation}\label{leg}
\begin{array}{rcl}
  \cal L_u &\colon &\sT N \supset N\times V^+\rightarrow \sT^\ast N \\
   &\colon &(x,v) \mapsto
\frac{m}{\<\tau,v\>}\imath_u^\ast\circ g\circ \imath_u(v)-
 \frac{m}{2\<\tau,v\>^2}\<g(\imath_u(v)),\imath_u(v)\>\tau -\varphi(x)\tau,
\end{array}
\end{equation}
    i.e. the vertical derivative of $\ell_{h,u}$ with respect to the projection $\sT
    N \rightarrow N$.

Since $\ell_{h,u}(x,v)-\ell_{h,u'}(x,v)=m\<\sigma(u',u),v\>$, we have also
\begin{equation} \label{leg2}
  \cal L_u(v) -\cal L_{u'}(v) = m\sigma(u',u).
\end{equation}

\begin{prop}
A mapping $\Phi_{u',u}\colon \mathsf T^\ast N \rightarrow \mathsf T^\ast N$
defined by
    \[\Phi_{u',u}(x,p)= (x, p + m\sigma(u',u))  \]
    has the following properties
\begin{enumerate}
  \item $\Phi_{u',u}(K_{m,u'})= K_{m,u}$,
  \item it is a symplectomorphism of the canonical symplectic structure  on
  $\mathsf T^\ast N$,
  \item $\mathsf T \Phi_{u',u}(D_{h,u'}) = D_{h,u}$.
\end{enumerate}
\end{prop}
\begin{pf} \label{leg3}
    The image of $\cal L_u$ is $K_{m,u}$, so the first property is an
    immediate consequence of  (\ref{leg2}) and the definition of $\Phi_{u',u}$.
    The mapping $\Phi_{u',u}$ is a translation by a constant vector. It
    follows that it is a symplectomorphism. Consequently,
    \[\sT \Phi_{u',u}((\sT K_{m,u'})^\S) = (\sT K_{m,u})^\S\] and
    \[\sT \Phi_{u',u}(\bar{D}_{h,u'} ) = \bar{D}_{h,u}.\]
    Since  $\Phi_{u',u}$ respects the time orientation, we have also
       \[\sT \Phi_{u',u}({D}_{h,u'} ) = {D}_{h,u}.\]
\end{pf}

 The above observation suggests the
following equivalence relation in $E_1\times N\times V^\ast$:

\begin{equation}
(u,x,p)\sim (u',x',p')\quad\Longleftrightarrow\quad\left\{
\begin{array}{l}
x=x',\\
p=p'+m\sigma(u',u).
\end{array}\right.
\end{equation}
Again, we have the obvious structure of the cartesian product in the set of
equivalence classes: $N\times P$. The set $N\times P$ will be called an {\it
affine phase space}.  The set $P$ is an affine space modelled on $V^\ast$:
$$[u,p]+\pi=[u, p+\pi]\text{ for }\pi\in V^\ast.$$
An element of $P$ will be denoted by $\p$.

It follows from Proposition~\ref{leg3} that $N\times P$ is a symplectic
manifold and the isomorphism of tangent and cotangent bundles assumes the
form
\begin{equation}\label{beta}
\begin{array}{rl}
 \beta &\colon \sT (N\times P)\simeq N\times P\times V\times
 V^\ast\longrightarrow \sT^\ast (N\times P)\simeq N\times P\times V^\ast\times
 V \\
    &\colon(x,\p,v,a) \longmapsto (x,\p, a, -v)
\end{array}
\end{equation}
 Moreover, the equivalence classes of the elements of mass-shells form the
universal mass shell $K_m$ and the elements of frame dependent dynamics form
the universal dynamics $D_h$  which is contained in $(\sT K_{m})^\S$.

    A straightforward calculation shows that  the function
        \[E_1\times N\times V^\ast\ni (u,x,p)\mapsto \frac{1}{2m}\<p,g'(p)\>+\<p,u\>\]
is constant on equivalence classes and projects to a function on
$N\times P$. We denote this function by $\Psi_m$. It follows that
the generating object (\ref{fam2}) of the dynamics $D_{h,u}$
defines a generating object
  \begin{equation} \label{fam3}\xymatrix@C=2cm{
N\times P\times \R_+ \ar[r]^-{-\widetilde{H}_{h}} \ar[d]_-{\zeta''}& \R\\
N\times V^\ast & },\end{equation}
    of the dynamics $D_h$, where
        \begin{equation}
\widetilde{H}_{h}(x,p,r)=r(\Psi_m+\varphi(x)).
\end{equation}
\subsection{Lagrangian as a generating object}
    In the previous section we have constructed the frame independent
dynamics $D_h$ and a hamiltonian generating object. Now, we show
that the frame independent affine lagrangian $\ell_h$ is also a
generating object of $D_h$. $\ell_h$ is a section of a bundle of
affine values $N\times W \rightarrow N\times V$ and its
differential is a section of $\sP (N\times W) \rightarrow N\times
V$ over $N\times V^+$. For a given frame $u$, we identify $N\times
W$ with $N\times V\times \R$ and a section of $\zeta$ with a
function on $N\times V$. Consequently, an affine  covector $a\in
\sP (N\times W)$ is represented by a covector $a_u\in
\sT^\ast(N\times V)= N\times V \times V^\ast \times V^\ast$. It
follows from (\ref{row}) that  $a_u= (x,v, a,b)$  and
$a_{u'}=(x,v,a,b +m\sigma(u,u'))$ represent the same element of
$\sP (N\times W)$. In the process of generation of frame dependent
dynamics we use the canonical isomorphism $\alpha_N  \colon
\sT\sT^\ast N\rightarrow \sT^\ast \sT N$ (\ref{alf}). We observe
that
    $$\alpha_N (x,a +m\sigma(u,u') ,v,b) = (x,v,b,a +m\sigma(u,u')), $$
hence $\alpha_N$ defines an isomorphism
    $$\alpha \colon \sT(N\times P)\rightarrow \sP(N\times W)$$
    and $(D_h)$ is the inverse image of $\d l_h(N\times V^+)$ by $\alpha$.

    Now, we can summarize our constructions. We have canonical symplectic
    structure on $N\times P$ with the corresponding mapping
$$\beta:\sT(N\times P)\longrightarrow\sT^\ast(N\times P),$$
    which forms the basis for the hamiltonian formulation of the dynamics.
    Together with $\alpha$ it gives rise to the following diagram (Tulczyjew
    triple):

$$\xymatrix@C-30pt{
(\sT^\ast(N\times P), \omega_{N\times P})\ar[dr] &
 &
(\sT (N\times P), d_{\sT}\omega_P) \ar[ll]_{\beta}\ar[rr]^{\alpha} \ar[dr]
\ar[dl] &
 &
(\sP(N\times W),\omega_{N\times W}) \ar[dl]\\
& N\times P & & N\times V & }
$$

\subsection{The Legendre transformation} \label{lt}

    The Legendre transformation is the passage from lagrangian to hamiltonian
generating object. In previous sections it was done with the
knowledge of the Legendre transformation for the frame dependent
dynamics. In that case we make use of the canonical
symplectomorphism $\gamma_M\colon \sT^\ast\sT^\ast M \rightarrow
\sT^\ast \sT M$ generated by $\langle \,,\,\rangle \colon \sT
M\times_M \sT^\ast M\rightarrow \mathbb{R}$, where $M$ is a
manifold and $\langle \,,\,\rangle$ is the canonical pairing
between vectors and covectors. It follows that the inverse image
$\gamma_M^{-1}(L)$ of a lagrangian submanifold $L$ generated by a
lagrangian $\ell$ is generated by a Morse family
    $$  \ell-\langle \,,\,\rangle \colon \sT M\times_M \sT^\ast M\rightarrow
    \mathbb{R},$$
    where $\sT M\times_M \sT^\ast M $ is considered a fibration over
    $\sT^\ast M$
    (see \cite{TU2} for details).

    Now, we show that analogous procedure can be applied in the case of
the affine framework.
    First, we observe that every element  $w\in W$ defines, in natural way,
an affine function on $P$:
\begin{equation}
f_w(\p)=\<p,v\>-r,\quad\text{where}\quad w=[u,v,r],\,\,\p=[u,p].
\end{equation}
Indeed, when we take another  representative of $w$ and $\p$, e.g.
$(u',v,r')$ and $(u',p')$ respectively, then we obtain
$$\<p',v\>-r'=\<p-m\sigma(u',u),v\>-r+m\<\sigma(u',u),v\>=\<p,v\>-r.$$ The element $w_1$
defines the constant function equal to $1$ on $P$:
$$f_{w_1}(p)=\<p,0\>-(-1)=1.$$
This implies the following:
\begin{prop}
There is a natural isomorphism between $P^\dag$ and $(W,w_1)$ given by
$$f_{[u,v,r]}([u,p])=\<p,v\>-r.$$
It means that also $W^{\ddag}\simeq P$.
\end{prop}
    With this isomorphism we have (see (\ref{dag})) $\sP(N\times W) \simeq N\times
    V\times V^\ast \times P$ and $\alpha \colon \sT(N\times P)\rightarrow
    \sP(N\times W)$ assumes the form
\begin{equation}\label{alph}
 \alpha \colon (x,\p,v, a)\longmapsto (x,v,a,\p).
\end{equation}

\noindent $(W,w_1)$, being a special vector space, has a structure of a
one-dimensional affine bundle modelled on $V\times\mathbf I$. The action
of the group $(\R, +)$ in the fiber over $V$ comes from the natural action
in the fiber of $E_1\times V\times \R\rightarrow E_1\times V$. The
fundamental vector field $X_W$ for this action is a constant vector field
with value $w_1$ at every point.

Now, we need a pairing  between $P$ and $V$, which reduces to $\langle
\,,\,\rangle$ (as a section of the trivial bundle $V\times V^\ast \times
\mathbb{R}$) in the vector case. The pairing is a section of $P\times W$ over
$P\times V$ defined by
\begin{equation}
P\times V\ni (\p,v)\longmapsto \<\p,v\>=[u,v,\<p,v\>]\in W,\quad
\text{where}\,\, \p=[u,p].
\end{equation}
The above definition is correct, i.e. does not depend on the choice of
representatives:
\begin{equation}\label{pair}
  [u,v,\<p,v\>]=[u',v,\<p,v\>-\<m\sigma(u',u),v\>]=[u',v,\<p',v\>].
\end{equation}

It remains to show that the pairing (\ref{pair}) generates an isomorphism
between $\sT (N\times P)$ and $\sP (V\times W)$.

\begin{prop}
There is a natural symplectomorphism between $\mathsf{P}((N\times
W)\times(N\times P))$ and $\mathsf{P}(N\times W)\ominus
\mathsf{T}^\ast(N\times P)$.
\end{prop}
    \begin{pf}
    It is enough to check that any section of $(N\times W)\times(N\times
P)$ over $(N\times V)\times(N\times P)$ is equivalent to a section $\sigma$
of the form
    $$\sigma(x,v,y,\p) =  \sigma_0(v) + f_1(x) - f_2(\p) -f_3(y), $$
where $\sigma_0$ is a linear section of $W\rightarrow V$ and functions $f_i$
are affine.
    \end{pf}

    Similar arguments show  that $\sP(N \times W)\simeq N\times V\times V^\ast
    \times P$.

    The canonical diagonal inclusion $N\subset N\times N$ implies the
projection
    $$V^\ast\times V^\ast \rightarrow V^\ast\colon (a,b)\mapsto a+ b$$
and consequently a relation between $\sP (N\times W \times P)$ and $\sP
((N\times W) \times (N\times P))$. With this relation a section of $\sP
(N\times W \times P)$ over $N\times V \times P$ defines a submanifold of
    $$\sP ((N\times W) \times (N\times P))= \sP(N\times W)\ominus \sT^\ast(N\times
    P)$$
i.e., a symplectic relation
    $$  \sT^\ast(N\times P)\longrightarrow\sP(N\times W).$$
    In particular, the differential of the pairing $\<\,,\,\>$ generates a
    relation

\begin{equation}
    \gamma \colon  \sT^\ast(N\times P)\simeq N\times P\times V^\ast \times
    V\longrightarrow\sP(N\times W) \simeq N\times V\times V^\ast\times P
\end{equation}
   It easy task to verify that this relation has the following representation
    $$N\times P\times V^\ast \times V \ni (x,\p,a, v) \longmapsto (x, -v, a, \p)
     \in N\times V\times V^\ast\times P.$$

    We see from (\ref{alph}) and (\ref{beta}) that $\gamma =\alpha\circ
    \beta^{-1}$, and  consequently $\gamma\circ\alpha(D_h) = \beta (D_h)$.
    Following the general rule for composing of generating objects (see \cite{LM}),  we
    conclude that $\beta (D_h)$ is generated by the Morse family

\begin{equation}
  \label{fam4}\xymatrix@C=2cm{
N\times P\times V^+ \ar[r]^-{-H_{h}} \ar[d]_-{\zeta''}& \R\\
N\times P&},
\end{equation}
    where
\begin{equation}
  H_h(x,v,\p) =  \< \p,v \> -\ell_h(x,v).
\end{equation}
    As in the frame-dependent case, this family can be reduced to the family
    (\ref{fam3}).

 \subsection{Inhomogeneous formulation of the dynamics}
    The inhomogeneous formulation of the dynamics is obtained by the reduction of the
    homogeneous one with respect to the canonical injection

        $$ \iota \colon W\supset W_1 \rightarrow W, $$
where  $W_1 = \zeta^{-1}(E_1)$ is an affine subspace of $W$.  The projection $\zeta$,
restricted to $W_1$, will be denoted by $\zeta_1$.
    $W_1$ is a special affine space modelled on a special vector space $W_0 = \zeta^{-1} (E_0)$.
    As in Section \ref{gav} we prove that
                        $$\sP W_1 = E_1\times W_0^\ddag $$
                        and consequently,
   $$\sP(N\times W_1) = N\times E_1 \times V^\ast \times W_0^\ddag . $$
   The affine space $W_0^\ddag$, which we denote by $P_0$,  is the quotient of
   $P=W^\ddag$ (Proposition 3) by the one-dimensional subspace of $V^\ast$,
   spanned by $\tau$. The canonical isomorphism
            $$\alpha \colon \sT(N\times P) = N\times P \times V \times V^\ast
            \rightarrow \sP(N\times W) = N \times V \times V^\ast \times P $$
(see Section \ref{lt}) projects to
 $$\alpha^1 \colon \sV^1(N\times P_0) = N\times P_0 \times E_1 \times E_0^\ast \rightarrow
 \sP_\sv(N\times W_1) = N \times E_1 \times E_0^\ast \times P_0, $$
 where $\sP_\sv$ is an affine analogue of $\sV^\ast$ (Section 3.1), i.e. $\sP_\sv (N\times W_1)$ contains affine differentials in vertical (with respect to the projection on $T$) directions only. We have obvious canonical projection
        $$\sP (N\times W_1)= N \times E_1 \times V^\ast \times P_0 \longrightarrow N \times E_1 \times E_0^\ast \times P_0 =\sP_\sv (N\times W_1) $$
    With the isomorphism $\alpha^1$ the restriction $\ell_i$ of $\ell_h$ to $\sV^1 N$
    generates a submanifold $D_i$ of $V^1(N\times P_0) \subset\sT (N\times P_0)$.

    The Hamiltonian formulation of the dynamics requires generalization of standard
    concepts like Poisson structure. The hamiltonian is, like lagrangian, not
    a function, but a section of a bundle of affine values.  Let us first notice
    that the pair $(P,\tau) $ is a special affine space, hence $N\times P$ is a
    bundle of affine values over $N\times P_0$. This is the bundle of a hamiltonian.
    The corresponding phase bundle $\sP (N\times P)$ is  canonically mapped
    to $\sT(N\times P_0) $.
(The detailed discussion of this mapping and of  the involved
geometric structures, like Lie affgebroid and affpoisson
structure, we postpone to a separate publication, see also
\cite{GGU}, \cite{PU}. The mathematical objects that can be used
for an inhomogeneous formulation of the dynamics are constructed
and discussed also in \cite{IMP}.)
    It follows that a section of $N\times P\rightarrow N\times P_0$
generates a vector  field on $N\times P_0$. For the dynamics of a
massive particle, the hamiltonian section is given by the equation
$\Psi_m+\varphi(x)=0 $ (Section \ref{ps}).

    \subsection{Affine Newtonian lagrangian bundle}
    As we have seen in \ref{fil}, the metric $g$ which has been used in the definition
    of the lagrangian appears (multiplied by the mass) as the Legendre map in the
    inhomogeneous, frame-dependent formulation of the dynamics. In  the frame-independent
    formulation, the Legendre map  is a mapping of the form
 \begin{equation}\label{legii}
\begin{array}{rcl}
  \cal L_{i} &\colon &\sV^1 N = N\times E_1 \rightarrow N\times P_0 \\
   &\colon &(x,w) \mapsto
     \overline{g},
\end{array}
\end{equation}
    where  $ \overline{g}$ is an affine mapping with the linear part $mg$.
     This observation justifies the following definition.
     For a special affine space $ \mathbf{A} = (A,v)$ with the model special vector
     space $(V,v)$, an \textit{affine metric} is a mapping
            $$h \colon \underline{A} \rightarrow V^\ddag$$
     such that its linear part $\underline{h}$ is a metric.
     The relation between affine metrics and kinetic energy part of lagrangians is
     established in  the following proposition.

    \begin{prop}
    Let $h \colon \underline{A} \rightarrow V^\ddag$  be an affine metric. There exist
    unique, up to a constant, section $\ell$ of the affine value bundle $A \rightarrow
    \underline{A}$ such that $\d \ell =h$.
    \end{prop}
    \begin{pf}
    Let $a$ be an arbitrary element in $A$ an let $ \underline{a}$ be its projection onto
    $ \underline{A}$. We define a section $\ell$ by the formula
    $$ \underline{A}\ni b \mapsto a + \langle h ( \underline{a} ), b- \underline{a} \rangle
    + \frac{1}{2} \langle\underline{h}(b- \underline{a}), b- \underline{a}\rangle,  $$
    where we interpret an element of $V^\ddag$ as a linear section of the bundle $V
    \rightarrow \underline{V}$,  and $\langle h (a), b- \underline{a} \rangle$ is the
    value of the section $ h (b)$ at the point $b- \underline{a}$.

    We have
    $$ \d \ell (b) =  h ( \underline{a} )  + \underline{h}(b- \underline{a}) = h(b)$$
    \end{pf}

\subsection{Concluding remarks}
    Frame independent inhomogeneous formulation of the dynamics requires affine bundles,
    affine values and an affine version of a metric tensor. The next step is to build
    a frame independent framework for  Hamilton-Jacobi theory and the wave mechanics
    (Schr\"odinger  equation).

\end{document}